\documentstyle[aps,preprint]{revtex}

\begin{document}

\tightenlines

\title{Flat Vacuum Branes Without Fine Tuning}

\author{Yu.~V.~Shtanov\footnote{E-mail: shtanov@bitp.kiev.ua}}
\address{Bogolyubov Institute for Theoretical Physics, Kiev 03143, Ukraine}

\maketitle

\bigskip

\begin{abstract} We construct solutions in purely gravitational bulk
with flat vacuum branes without fine tuning the constants of the theory.  Our
solutions incorporate one or two branes and have topology $S^1 \times R^4$
without $Z_2$ symmetry. They are not invariant under the four-dimensional
Lorentz group effectively acting on the brane, but the significance of this
fact will become clear only after the study of their perturbations.
\end{abstract}


There is some hope that the long-standing problem of cosmological constant can
be alleviated in the context of braneworld theories that involve extra
dimensions. The idea of this kind was put forward already in \cite{RSh} in a
(4+2)-dimensional theory with the warped metric ansatz. Recently, it was
revived in the context of (4+1)-dimensional models with branes. As regards the
simplest such models, it is known that the existence of solutions with flat
vacuum branes requires fine tuning the bulk cosmological constant and brane
tension, as it is the case in the Randall--Sundrum original model \cite{RS}. In
an attempt to avoid fine tuning, one usually employs additional fields
propagating in the bulk; for example, one considers self-tuning mechanisms
based on the dynamics of bulk scalar fields or antisymmetric tensor field;
however, this approach is still far from being satisfactory (see \cite{CC} for
recent reviews).

In most of the approaches to this problem, one looks for solutions in the bulk
that have a $Z_2$ symmetry of reflection with respect to the brane and are
invariant under the four-dimensional Poincar\'{e} group effectively acting on
the brane.  Solutions incorporating flat vacuum branes with $Z_2$ symmetry of
reflection but without four-dimensional Lorentz symmetry of the bulk were
recently considered in \cite{CEG}. In this case, the nongravitational physics
confined to the brane remains to be Lorentz invariant, but bulk gravitational
effects might cause apparent violation of Lorentz invariance from the viewpoint
of an observer on the brane.  The fine-tuning problem of the cosmological
constant of the brane is alleviated in \cite{CEG} by the introduction of an
additional vector field in the bulk. Meanwhile, {\em cosmological\/} braneworld
solutions without $Z_2$ symmetry were thoroughly studied in a series of papers
\cite{nonsym}.

In this paper, we show that, without imposing $Z_2$ symmetry on the bulk, it is
possible to construct solutions with flat vacuum branes in the simplest purely
gravitational model without fine tuning the constants of the theory. The
absence of $Z_2$ symmetry in this case implies also the absence of the
four-dimensional Lorentz symmetry of solution in the bulk. However, similarly
to the situation considered in \cite{CEG}, the nongravitational field theory
confined to the brane remains to be Lorentz symmetry, and the four-dimensional
translation symmetry of the bulk is also preserved. The significance of the
fact that the five-dimensional gravitational background is not Lorentz
invariant must become clear only after the study of its perturbations.

We consider the theory with the action
\begin{equation} \label{action}
S = M^3 \sum \left[ \int_{\mathrm bulk} \left({\mathcal R} - 2 \Lambda
\right) - 2 \int_{{\mathrm brane}} K  \right] + \int_{{\mathrm brane}} L
(h_{ab}, \phi) \, ,
\end{equation}
where ${\mathcal R}$ is the scalar curvature of the metric in the bulk, $\phi$
is the collection of fields confined to the brane, $h_{ab}$ is the induced
metric on the brane, and $K = h^{ab}K_{ab}$ is the trace of the tensor of its
extrinsic curvature, which is defined as $K_{ab} = h^c{}_a \nabla_c n_b$ with
$n^a$ being the inner normal in the corresponding volume. The sum in
(\ref{action}) is taken over all bulk components bounded by branes, and, for
simplicity, we assume that all bulk components have one and the same
five-dimensional Planck mass $M$ and cosmological constant $\Lambda$. The
Lagrangian $L (h_{ab}, \phi)$ describes the matter restricted to the brane and
can include the purely gravitational terms constructed from the induced metric
$h_{ab}$ on the brane.  A brane which is embedded in a five-dimensional bulk is
described by the junction equation
\begin{equation} \label{brane}
- M^3 \left[ S^{(+)}_{ab} + S^{(-)}_{ab} \right] = \tau_{ab} \, ,
\end{equation}
where $\tau_{ab}$ is the effective stress--energy tensor of the brane, which
stems from the variation of the last term in (\ref{action}) with respect to the
induced metric, and the tensors $S^{(\pm)}_{ab}$ are expressed through the
extrinsic curvatures $K^{(\pm)}_{ab}$ of the brane in the corresponding two
volumes according to the definition
\begin{equation}
S^{(\pm)}_{ab} = K^{(\pm)}_{ab} - h_{ab} K^{(\pm)} \, , \quad K^{(\pm)} =
h^{ab} K^{(\pm)}_{ab} \, .
\end{equation}

In the case under consideration, the constraint equations on the brane are
equivalent to the
following system of equations \cite{Shtanov}:
\begin{equation} \label{constraint}
R - 2\Lambda + \frac14 \left(S_{ab} S^{ab} - \frac13 S^2 \right) + \frac14
\left(Q_{ab} Q^{ab} - \frac13 Q^2 \right) = 0 \, ,
\end{equation} \begin{equation} \label{orthog}
S_{ab} Q^{ab} - \frac13 S Q = 0 \, ,
\end{equation} \begin{equation} \label{con}
D_a Q^a{}_b = 0 \, ,
\end{equation}
where $R$ is the scalar curvature of the induced metric on the brane,
$S_{ab} = S^{(+)}_{ab} + S^{(-)}_{ab}$, $Q_{ab} = S^{(+)}_{ab} -
S^{(-)}_{ab}$, $S = h^{ab} S_{ab}$, $Q = h^{ab} Q_{ab}$, and $D_a$ denotes
the covariant derivative on the brane associated with the induced metric
$h_{ab}$. Imposing the standard $Z_2$ orbifold symmetry is equivalent to
setting $Q_{ab} = 0$.

Our procedure consists in obtaining a solution of the system of equations
(\ref{brane}), (\ref{constraint})--(\ref{con}) on the brane and then
matching it with solution in the bulk.  Since we are looking for flat vacuum
solutions on the brane, we have from Eq.~(\ref{brane}):
\begin{equation}\label{flat}
S^a{}_b = {\sigma \over M^3} h^a{}_b \, ,
\end{equation}
where $\sigma$ is the vacuum brane tension.\footnote{Equation of the form
(\ref{flat}) is valid in general whenever the stress--energy tensor on the
brane is proportional to the metric tensor, which is true in our case.} Then
Eq.~(\ref{orthog}) gives the condition $Q = 0$. Thus, the extrinsic curvatures
of the two sides of the brane are given by
\begin{equation}\label{data}
K_{(\pm)}{}^a{}_b = - {\sigma \over 6 M^3} h^a{}_b \pm \frac12 Q^a{}_b \, ,
\end{equation}
where $Q^a{}_b$ is a conserved traceless tensor on the brane. The constraint
equation (\ref{constraint}) leads to the following condition:
\begin{equation} \label{tuning}
2 \Lambda + {\sigma^2 \over 3 M^6} - q = 0  \, ,
\end{equation}
where
\begin{equation}
q = \frac14 Q^{ab} Q_{ab} \, .
\end{equation}

In passing, we note that imposing $Z_2$ symmetry would imply $q = 0$, and
Eq.~(\ref{tuning}) would then turn to the well-known fine-tuning condition
\cite{RS} between the constants of the theory.  Without imposing $Z_2$
symmetry, we have a free parameter $q$ at our disposal, which makes fine
tuning unnecessary.  The price that we pay for this is that solution in the
bulk is no longer invariant under the four-dimensional Lorentz group
effectively acting on the brane.

The data (\ref{data}) on the flat brane must be matched with the solution of
the Einstein equations in the bulk.  To solve this problem, we choose the
Gaussian normal coordinates in the bulk generated by the Minkowski
coordinates $x^\alpha$, $\alpha = 0,\ldots,3$, on the brane, so that the
metric of the solution in these coordinates has the form
\begin{equation}
d s^2_5 = dy^2 + h_{\alpha\beta} (y, x) dx^\alpha dx^\beta \, ,
\end{equation}
where $y$ is the fifth coordinate in the bulk, and the brane corresponds to $y
= 0$. One can obtain the differential equations for the components
$h_{\alpha\beta}$ and $K_{(\pm)}{}^\alpha{}_\beta$, in the range $y \ge 0$ for
``$+$'', and $y \le 0$ for ``$-$'', by using the procedure of $4+1$ splitting
of the Einstein equations with the following result (see \cite{Shtanov}):
\begin{equation} \label{s}
\pm {\partial K_{(\pm)}{}^\alpha{}_\beta \over \partial y} = R^\alpha{}_\beta -
\frac23 \Lambda \delta^\alpha{}_\beta - K_{(\pm)} K_{(\pm)}{}^\alpha{}_\beta
\, ,
\end{equation} \begin{equation} \label{metric}
\pm {\partial h_{\alpha\beta} \over \partial y} = 2 h_{\alpha\gamma}
K_{(\pm)}{}^\gamma{}_\beta \, ,
\end{equation}
where $R^\alpha{}_\beta$ are the components of the Ricci tensor of the metric
$h_{\alpha\beta}$.  Together with the constraint equations
(\ref{constraint})--(\ref{con}), Eqs.~(\ref{s}) and (\ref{metric}) are
equivalent to the Einstein equations in the bulk. The initial data for these
equations at $y = 0$ for the domains $y \ge 0$ and $y \le 0$ is given by
Eq.~(\ref{data}) and by the condition $h_{\alpha\beta} |_{y = 0} =
\eta_{\alpha\beta}$.

The components $Q^\alpha{}_\beta$ of the traceless conserved tensor $Q^a{}_b$
can be chosen constant on the brane. Then, because of the homogeneity of the
data on the brane, the solution for $h_{\alpha\beta}$ and
$K_{(\pm)}{}^\alpha{}_\beta$ depends only on $y$; thus, the induced metric
$h_{\alpha\beta}$ is flat on every hypersurface $y = {\mathrm const}$.
Consequently, the components of the Ricci tensor $R^\alpha{}_\beta$ in
Eq.~(\ref{s}) are identically equal to zero, and this equation becomes closed
with respect to $K_{(\pm)}{}^\alpha{}_\beta (y)$. Its solution can be presented
in the form
\begin{equation} \label{sol-s}
K_{(\pm)}{}^\alpha{}_\beta (y) = \frac12 \Bigl[ - A \left(|y|\right)
\delta^\alpha{}_\beta \pm B \left(|y|\right) Q^\alpha{}_\beta \Bigr] \, ,
\end{equation}
where the functions $A(y)$ and $B(y)$ are defined for $y \ge 0$. These
functions satisfy the constraint equation that follows from
Eq.~(\ref{constraint}):
\begin{equation} \label{constr}
2\Lambda + 3 A^2 - q B^2 \equiv 0 \, ,
\end{equation}
and the system of ordinary differential equations that follows from (\ref{s}):
\begin{equation} \label{diff}
{d A \over d y} = \frac43 \Lambda + 2 A^2 \, , \quad {d B \over d y} = 2 A B
\, ,
\end{equation}
with the initial conditions
\begin{equation} \label{ini}
A(0) = {\sigma \over 3 M^3} \, , \quad B(0) = 1 \, .
\end{equation}
Solution of (\ref{diff}), (\ref{ini}) exists in some interval of $y > 0$ and
can be expressed in terms of elementary functions. If the values of $q$ and
$\sigma/M^3$ are positive, then both functions $A(y)$ and $B(y)$ are positive
and monotonically increase to infinity on a finite interval of $y$. After one
obtains the solution for $A(y)$ and $B(y)$, one can solve Eq.~(\ref{metric})
for $h_{\alpha\beta} (y)$ with the initial condition $h_{\alpha\beta} (0) =
\eta_{\alpha\beta}$, and the corresponding solution can also be expressed in
terms of elementary functions. We will not write these solutions here. We only
note that, if there exist Minkowski coordinates on the brane in which the
matrix of the tensor components $Q^\alpha{}_\beta$ is diagonal and invariant
under spatial rotations, then the solution in the bulk reproduces the
Schwarzschild--(anti)~De~Sitter space expressed in Gaussian normal coordinates.
Their connection with the usual spherical coordinates was under investigation
in \cite{MSM}.

To give an example of complete solution in the bulk, we choose the traceless
matrix of components $Q^\alpha{}_\beta$ to be diagonal. Then solution
(\ref{sol-s}) and, consequently, solution $h_{\alpha\beta} (y)$ of
Eq.~(\ref{metric}) also have diagonal form.  Let the solution of our equations
exist in a segment $[-y_0, y_0]$ of $y$. Since the induced metrics on the
hypersurfaces $y = {\mathrm const}$ are flat, there exists an isometry between
the two hypersurfaces $y = \pm y_0$, which we can use to identify them. To
define this isometry, it suffices to perform certain uniform dilatations of the
coordinates $x^\alpha$ of one of the hypersurfaces, say, $y = y_0$, which will
modify the components of the induced metric $h_{\alpha\beta} (y_0)$ and make
them equal to the components $h_{\alpha\beta} (-y_0)$. Note that such
dilatations will not change the components $K_{(+)}{}^\alpha{}_\beta (y_0)$
because they form a diagonal matrix. After this, we place another brane at the
hypersurface obtained by the identification described. Since the junction
condition (\ref{brane}) must be satisfied on this new brane, its tension
$\sigma_0$ must satisfy the equation
\begin{equation} \label{junction}
M^3 \left[ S_{(+)}{}^\alpha{}_\beta (y_0) + S_{(-)}{}^\alpha{}_\beta (-y_0)
\right] = 3 M^3 A\left( y_0 \right) \delta^\alpha{}_\beta = - \sigma_0
\delta^\alpha{}_\beta\, ,
\end{equation}
i.e.,
\begin{equation} \label{tension}
\sigma_0  = - 3 M^3 A\left( y_0 \right) \, .
\end{equation}
If $\sigma/M^3$ is positive, then $\sigma_0/M^3$ must necessarily be negative
and greater by absolute value because $A(y)$ is monotonically increasing with
$y$ in the present case.

It is also possible to construct solutions with a single brane and with
topology $S^1 \times R^4$.  To do this, we again choose the traceless matrix of
components $Q^\alpha{}_\beta$ to be diagonal. We then impose the following
initial conditions for the functions $A(y)$ and $B(y)$:
\begin{equation} \label{zero}
A(0) = 0 \, , \quad B(0) = 1 \, ,
\end{equation}
so that the extrinsic curvature is continuous and no brane is present at $y =
0$.  Since the value of $q$ is positive in the case under consideration, the
constraint equation (\ref{constr}) implies that the value of the bulk
cosmological constant $\Lambda$ must be positive.  By solving the system of
differential equations (\ref{diff}), we obtain solution in the form
(\ref{sol-s}) and then solve for the metric components $h_{\alpha\beta} (y)$.
Since solution exists in some segment $[-y_0, y_0]$ of $y$, we can perform
identification of the two flat hypersurfaces $y = \pm y_0$ as we did before. At
this hypersurface of identification, a flat vacuum brane is to be inserted. It
is easy to see from the first of equations (\ref{diff}) that the sign of
$A(y_0)$ in the present case will coincide with the sign of $\Lambda$.  Thus,
the brane tension will be given by Eq.~(\ref{tension}) and will be negative in
the present case.

We describe a class of such single-brane solutions in more detail.  We choose
solution in the bulk with positive $\Lambda$ in the form
\begin{equation}\label{wall}
ds^2 = - f(r) dt^2 + {dr^2 \over f(r)} + {r^2 \over \ell^2} \sum_{i=1}^3 \left(
dx^i \right)^2 \, ,
\end{equation}
where
\begin{equation}
f(r) = {r_g^2 \over r^2} - {r^2 \over \ell^2} \, , \quad \ell = \sqrt{6 \over
\Lambda} \, ,
\end{equation}
and $r_g$ is an arbitrary positive constant. We consider this solution in the
domain $f(r) > 0$, which corresponds to $0 < r < \sqrt{r_g \ell}\,$. The
hypersurfaces $r = {\mathrm const}$ have flat induced metrics on them.  The
coordinate $r$ is related to the Gaussian normal coordinate $y$ by
\begin{equation}
y = \int_{r_0}^r {dr' \over \sqrt{f(r')}} \, .
\end{equation}

The regular (no brane) hypersurface $r = r_0$ corresponding to $y =0$ is
determined by condition (\ref{zero}), which requires that the trace of the
tensor of extrinsic curvature of this hypersurface be zero.  Calculating the
tensor of extrinsic curvature $K^\alpha{}_\beta (r)$ of the hypersurface $r =
{\mathrm const}$ with the unit normal in the direction of increasing $r$, we
obtain
\begin{equation}
K^\alpha{}_\beta (r) = {\mathrm diag} \left\{ - {1 \over \sqrt{f(r)}} \left(
{r_g^2 \over r^3} + {r \over \ell^2} \right),\ {\sqrt{f(r)} \over r}\, ,\
{\sqrt{f(r)} \over r}\, ,\ {\sqrt{f(r)} \over r} \right\} \, ,
\end{equation}
so that
\begin{equation}
K(r) = - {1 \over \sqrt{f(r)}} \left( {r_g^2 \over r^3} + {r \over \ell^2}
\right) + {3 \sqrt{f(r)} \over r}
\end{equation}
and
\begin{equation}\label{asym}
Q^\alpha{}_\beta (r) \equiv 2 K^\alpha{}_\beta (r) - \frac12 K(r)
\delta^\alpha{}_\beta = g(r)\, {\mathrm diag} \{ -3, 1, 1, 1 \} \, ,
\end{equation}
where
\begin{equation}\label{g}
g(r) = \frac12 \left[ {1 \over \sqrt{f(r)}} \left( {r_g^2 \over r^3} + {r \over
\ell^2} \right) + {\sqrt{f(r)} \over r} \right] \, .
\end{equation}
Thus, we have the equation for $r_0$:
\begin{equation}
K(r_0) = 0 \quad \Longrightarrow \quad r_0 = \left( {r_g \ell \over \sqrt{2}}
\right)^{1/2} \, .
\end{equation}

It is easy to verify that the function $g(r)$ in Eq.~(\ref{asym}) defined by
(\ref{g}) reaches its minimum $g(r_0) = 2 / \ell\,$ at $r = r_0$, in perfect
agreement with the differential equations (\ref{diff}) and conditions
(\ref{zero}).  It remains to shift from the hypersurface $r = r_0$ to the right
and to the left so that the tensor components $Q^\alpha{}_\beta (r)$ have
identical values on the two new hypersurfaces. We obtain the condition
\begin{equation}\label{shift}
g(r_1) = g(r_2)
\end{equation}
to be satisfied for $0 < r_1 < r_0 < r_2 < \sqrt{r_g \ell}$. This equation has
infinitely many solutions; specifically, for every $r_1$ in the range $0 < r_1
< r_0$, there exists a unique $r_2$ in the range $r_0 < r_2 < \sqrt{r_g \ell}$
for which (\ref{shift}) is satisfied, and vice versa.  This is a consequence of
the fact that the function $g(r)$ monotonically decreases from positive
infinity to its minimum value $2 / \ell\,$ as $r$ increases from $0$ to $r_0$,
and then monotonically increases from $2 / \ell\,$ to positive infinity as $r$
increases from $r_0$ to its limiting value $\sqrt{r_g \ell}\,$.  The values $r
= r_1$ and $r = r_2$ correspond to the values $y = - y_0$ and $y = y_0$,
respectively, of the Gaussian normal coordinate.

Choosing an arbitrary solution $(r_1, r_2)$ of Eq.~(\ref{shift}), we
isometrically identify the corresponding flat hypersurfaces and insert the
vacuum brane at this position. The junction condition (\ref{junction}) then
reads
\begin{equation} \label{tension1}
\frac34 M^3 \left[ K(r_2) - K(r_1) \right] = 3 M^3 \left( {\sqrt{f(r_2)} \over
r_2} - {\sqrt{f(r_1)} \over r_1} \right) = \sigma_0 \, .
\end{equation}

The identification of hypersurfaces can be described as follows:  If
$\left(t_s, x_s^i\right)$ are the coordinates on the hypersurfaces $r = r_s$,
$s = 1,2$, in the metric (\ref{wall}), then the identification formulas are
written as
\begin{equation}\label{identify}
\epsilon \left( t_1 - t_0 \right) =  \sqrt{f(r_2) \over f(r_1)} \, t_2 \, ,
\quad \sum_j O^i{}_j \left( x_1^j - x_0^j \right) = {r_2 \over r_1} \, x_2^i \,
,
\end{equation}
where the parameter $\epsilon = \pm 1$ and the orthogonal matrix $O^i{}_j$
reflect the possibility of identifying the two hypersurfaces with different
time and space orientations (a procedure quite similar to the construction of
the M\"obius strip) and with spatial rotation, and $t_0$ and $x_0^i$, $i =
1,2,3$, are arbitrary parameters corresponding to the coordinates on the
hypersurface $r = r_1$ which are identified with the origin of coordinates on
the hypersurface $r = r_2$. The identification parameters have the following
geometrical meaning: the spacelike geodesic normal to the brane proceeding from
the point with natural Minkowski coordinates $\left(T, X^i \right)$ on the
brane in the direction of decreasing $r$ goes around the compact fifth
dimension and hits the brane at the point with Minkowski coordinates
\begin{equation}
T_* = \epsilon  {\sqrt{f(r_1) \over f(r_2)}} \left( T  - T_0 \right) \, , \quad
X^i_* = {r_1 \over r_2} \sum_j O^i{}_j \left(X^j - X_0^j \right) \, ,
\end{equation}
where $T_0 = \sqrt{f(r_2)}\, t_0$ and $X_0^j = x_0^j r_2 / \ell$.

Thus, the constructed solution is described by metric (\ref{wall}) in the range
$r_1 \le r \le r_2$ with the isometric identification of the hypersurfaces $r =
r_1$ and $r = r_2$ described by (\ref{identify}) and vacuum brane located at
this position with tension given by (\ref{tension1}).  The solution is
obviously nonsingular: every geodesic either is infinitely extendible or hits
the brane. The solutions with two branes described above have similar
properties and can be analyzed in a similar way.

We conclude that we constructed regular solutions in the bulk with one or two
flat branes without fine tuning the constants of the theory. Conditions
(\ref{tension}) and (\ref{tension1}) for $\sigma_0$ do not represent strict
constraints since the point $y_0$ can be chosen arbitrarily in the domain of
existence of the solution of Eqs.~(\ref{diff}) and (\ref{metric}). Our
solutions have topology $S^1 \times R^4$, as in the Randall--Sundrum model
\cite{RS}, but without $Z_2$ symmetry. They are not invariant under the
four-dimensional Lorentz group effectively acting on the brane, although they
remain to be invariant with respect to the four-dimensional group of
translations.  The largest natural symmetry that can persist in the bulk is the
three-dimensional spatial symmetry of Euclidean motions with a distinguished
time direction. The significance of this fact will become clear after the study
of perturbations around the background obtained. We note, however, that the
nongravitational physics on the branes remains to be Poincar\'{e} invariant.

\bigskip

This research was supported in part by the INTAS grant for project
No.~2000-334.


\begin{thebibliography}{99}

\bibitem{RSh}
V.~A.~Rubakov and M.~E.~Shaposhnikov, Phys.\@ Lett.\@ B {\bf 125}, 139 (1983).

\bibitem{RS}
L.~Randall and R.~Sundrum, Phys.\@ Rev.\@ Lett.\@ {\bf 83}, 3370 (1999)
[{\tt hep-ph/9905221}];\ Phys.\@ Rev.\@ Lett.\@ {\bf 83}, 4690 (1999) [{\tt
hep-th/9906064}].


\bibitem{CC}
C.~Cs\'aki, J.~Erlich, C.~Grojean, and T.~J.~Hollowood, Nucl.\@ Phys.\@ B
{\bf 584}, 359 (2000) [{\tt hep-th/0004133}];\ H.~P.~Nilles, {\sl The
cosmological constant and the brane world scena\-rio\/}, {\tt
hep-ph/0101015};\ J.~E.~Kim, {\sl Self-tuning solutions of the cosmological
constant\/}, {\tt hep-th/0108081}.

\bibitem{CEG}
C.~Cs\'aki, J.~Erlich, and C.~Grojean, Nucl.\@ Phys.\@ B {\bf 604}, 312 (2001)
[{\tt hep-th/0012143}];\ C.~Cs\'aki, J.~Erlich, and C.~Grojean, {\sl The
cosmological constant problem in brane-worlds and gravitational Lorentz
violations\/}, {\tt gr-qc/0105114}.

\bibitem{nonsym}
P.~Kraus, JHEP {\bf 9912}, 011 (1999) [{\tt hep-th/9910149}]; \ D.~Ida, JHEP
{\bf 0009}, 014 (2000) [{\tt gr-qc/9912002}]; \ H.~Collins and B.~Holdom,
Phys.\@ Rev.\@ D {\bf 62}, 105009 (2000) [{\tt hep-ph/0003173}]; \ N.~Deruelle
and T.~Dole{\v z}el, Phys.\@ Rev.\@ D {\bf 62}, 103502 (2000) [{\tt
gr-qc/0004021}]; \ H.~Stoica, H.~Tye, and I.~Wasserman, Phys.\@ Lett.\@ B {\bf
482}, 205 (2000) [{\tt hep-th/0004126}]; \ P.~Bowcock, C.~Charmousis, and
R.~Gregory, Class.\@ Q.\@ Grav.\@ {\bf 17}, 4745 (2000) [{\tt hep-th/0007177}];
\ A.-C.~Davis, I.~Vernon, S.~C.~Davis, and W.~Perkins, Phys.\@ Lett.\@ B {\bf
504}, 254 (2001) [{\tt hep-ph/0008132}]; \ W.~B.~Perkins, Phys.\@ Lett.\@ B
{\bf 504}, 28 (2001) [{\tt gr-qc/0010053}]; \ B.~Carter and J.-P.~Uzan, Nucl.\@
Phys.\@ B {\bf 406}, 45 (2001) [{\tt gr-qc/0101010}]; \ R.~A.~Battye,
B.~Carter, A.~Mennim, and J.-P.~Uzan, Phys.\@ Rev.\@ D {\bf 64}, 124007 (2001)
[{\tt hep-th/0105091}].

\bibitem{Shtanov}
Yu.~V.~Shtanov, {\sl Closed system of equations on a brane\/}, {\tt
hep-ph/0108153}.

\bibitem{MSM}
S.~Mukohyama, T.~Shiromizu, and K.~Maeda, Phys.\@ Rev.\@ D {\bf 62},
024028 (2000); Erratum---Phys.\@ Rev.\@ D {\bf 63} 029901 (2001)
[{\tt hep-th/9912287}].

\end{thebibliography}
\end{document}